\documentclass{aa}

\usepackage{graphicx}
\usepackage{txfonts}
\usepackage{color}
\usepackage{multirow}
\usepackage{bm}

\def\kmsmpc{\rm \,km\,s^{-1}\,Mpc^{-1}}
\def\kms{\rm \,km\,s^{-1}}
\def\age{\rm age}
\def\form{\rm form}
\def\Univ{\rm U}
\def\RL{R_{\rm L}}
\def\msun{M_{\sun}}

\titlerunning {Environmental Effects on LRGs as Cosmic Chronometers}
\authorrunning {Liu et al.}

\begin{document} 

\title{Quiescent luminous red galaxies  as cosmic chronometers:
on the significance of mass and environmental dependence}

\subtitle{} 
\author{G.\ C. Liu\inst{1,6} \and Y.\ J. Lu\inst{2} \and L.\ Z. Xie\inst{3,4}
\and X.\ L. Chen\inst{2,5} \and Y.\ H. Zhao\inst{2}
}

\institute{College of Science, China Three Gorges University, Yichang
443002, China 
\and
Key Laboratory of Optical Astronomy, National Astronomical
Observatories, Chinese Academy of Science, Beijing 100012, China
\email{luyj@nao.cas.cn}  
\and
National Astronomical Observatories, Chinese Academy of Science,
Beijing 100012, China 
\and
Istituto Nazionale di Astrofisica (INAF), Osservatorio Astronomico di
Trieste, Via Tiepolo 11, I-34131 Trieste, Italy 
\and
Center of High Energy Physics, Peking University, Beijing 100871,
China 
\and
Department of Astronomy and Astrophysics, Peking University, Beijing 100871,
China 
}

\date{Received ; Accepted }
 
\abstract
%context heading (optional)
{ Massive luminous red galaxies (LRGs) are believed to be evolving
passively and can be used as cosmic chronometers to estimate the
Hubble constant (the differential age method). However, different LRGs
may be located in different environments. The environmental effects, if
any, on the mean ages of LRGs, and the ages of the oldest LRGs at
different redshift, may limit the use of the LRGs as cosmic
chronometers. }
%aims heading (mandatory)
{We aim to investigate the environmental and mass dependence of the
formation of `quiescent' LRGs, selected from the Sloan Digital Sky
Survey Date Release 8 (SDSS), and to pave the way for using  LRGs as cosmic
chronometers. }
{Using the population synthesis software STARLIGHT, we derive the
stellar populations in each LRG through the full spectrum fitting and
obtain the mean age distribution and the mean star formation history
(SFH) of those LRGs.  }
{We find that there is no apparent dependence of the mean age and the
SFH of quiescent LRGs on their environment, while the ages of those
quiescent LRGs  depend weakly on their mass.  We compare the SFHs of
the SDSS LRGs with those obtained from a semi-analytical galaxy
formation model and find that they are roughly consistent with each
other if we consider the errors in the STARLIGHT-derived ages.  We
find that a small fraction of later star formation in LRGs leads to a
systematical overestimation ($\sim 28\%$) of the Hubble constant by
the differential age method, and the systematical errors in the
STARLIGHT-derived ages may lead to an underestimation $(\sim 16\%)$ of
the Hubble constant.  However, these errors can be corrected by a
detailed study of the mean SFH of those LRGs and by calibrating the
STARLIGHT-derived ages with those obtained independently by other
methods.  }
{The environmental effects do not play a significant role in the age estimates of
quiescent LRGs; and the quiescent LRGs as a population can be
 used securely as cosmic chronometers, and the Hubble constant can be
measured with high precision by using the differential age method.}


\maketitle

\section{Introduction}
Luminous red galaxies (LRGs) are believed to   evolve passively and
host the oldest stellar population, and  can thus be used as cosmic
chronometers to measure the Hubble constants at different redshifts
(the differential age method; \citealt{Jim02}). For example,
\citet[][]{Jim03} adopted the differential age method to measure
the Hubble constant at redshift $z\sim 0$ by using the oldest LRGs at
different redshifts among a large number of LRGs from the Sloan
Digital Sky Survey (SDSS). The differential age method is also adopted
to measure the Hubble constants at higher redshifts by using red
galaxies from other surveys, and those measurements were used to
constrain other cosmological parameters \citep[e.g.,][]{Simon05,
Stern10, Moresco12, Moresco15}. 
\citet{Liu12} further developed the differential age method by using
the mean ages of the SDSS LRGs, rather than the oldest LRGs, to
constrain the Hubble constant at low redshift, for which a homogeneous
sample of LRGs is required.  However, different LRGs may be located in
different environments, for example, some LRGs are the brightest
central galaxies (BCGs) or satellite galaxies of clusters located in
dense environments, and some LRGs are field galaxies located in less
dense environments.  Therefore, the environmental effects may affect
the accuracy of the measurements of both the mean ages of LRGs and
the ages of the oldest LRGs at different redshift (especially when
they are not selected from a single uniform survey) and  may thus limit
the use of the LRGs as cosmic chronometers. Furthermore, the later
star formation in the LRGs, if any, may also introduce systematic
errors to the use of LRGs as  cosmic chronometers (see a discussion
in \citealt{Liu12}).  These effects must be addressed, not only for
the validity of the use of LRGs as cosmic chronometers, but also to
improve the accuracy of the estimates of the Hubble constant and other
cosmological parameters via the differential age method.
 
The environmental effects on galaxy formation have been studied
extensively in the past decade \citep[e.g.,][]{For98, Bla05, Coo06,
Luc06, Pog06, Lee10, Tho10}. The differences in the assembly histories
of the dark matter halos of those galaxies in different environments
appear to be the main driver of the environmental effects detailed
below.  Galaxies in high density environments may experience many more
mergers, leading to multiple starburst episodes in their formation
and evolution histories, compared to those in low density
environments \citep[e.g.,][]{Tho05}; The BCGs in massive dark matter
halos may have formed at early time and  were then quenched because of internal quenching processes, such as the AGN feedback
\citep{Luc07}; the hot gas in satellite galaxies may have been removed due to
ram pressure stripping (an external process) when these galaxies fell
into a group or cluster of galaxies, and thus the star formation in
them was suppressed \citep{Gom03, Bal04a, Pas10, Luc12}; dynamical
perturbations on those infalling galaxies may also have led to some
instabilities and transformed disk galaxies to spheroidal galaxies
\citep{Naa09}. These differences in the formation histories of BCGs,
satellite galaxies and field galaxies may be imprinted on their
structure properties, as studied extensively over many decades
\citep{Mat64, Oem73, Sch88, Gra96, Pat06, Ber07, Des07, Lau07, Lin07,
Luc07, Liu08}.

In this paper, we adopt the spectral population synthesis technique to
study the mass and environmental effects of the so-called quiescent SDSS LRGs
and their star formation histories (SFHs), and  check whether
they can be securely used as cosmic chronometers to measure the Hubble
constants at different redshifts. 
Several previous studies investigated the ages of early-type galaxies
as a function of their environment and masses (Bernardi et al.1998, 2006,
Thomas et al. 2005, 2010). The present paper differs from  previous
studies in several aspects: (1) we adopt the largest LRG (or early type
galaxy) sample (7882); (2) we use the full spectrum fitting method to 
extract galaxy properties; (3) we investigate the detailed SFHs for 
different type of early type galaxies, i.e., BCGs, member galaxies, and field galaxies.
 In Section~\ref{sec:sample}, we
describe the selection criteria for `quiescent' LRGs in different
environments, e.g., the central and satellite galaxies in clusters of
galaxies and dense environments, and field galaxies in less dense
environments.  In Section~\ref{sec:spectralfit}, we adopt the spectral
population synthesis technique and the full spectrum fitting method to
obtain the SFHs and age properties of these LRGs. In Section~\ref{sec:results}, we investigate the
mass and environmental dependence of the formation of LRGs by
comparing the obtained SFHs and age properties of LRGs with different
mass and in different environments.
In
Section~\ref{sec:compsammodel}, we further investigate the underlying reasons for the little
environmental effect in the formation of LRGs  by comparing the SFHs and the age
properties of the SDSS LRGs with mock LRGs from a semi-analytical
galaxy formation model \citep{Guo11}. 
We analyze the possible uncertainties in the estimates of the Hubble
constant via the differential age method in
Section~\ref{sec:discussion}.
No significant environmental dependence of the age properties of
quiescent LRGs suggests that these LRGs can be securely used as cosmic
chronometers. Conclusions are given in Section~\ref{sec:conclusion}.

%__________________________________________________________________

\section{Sample selection}\label{sec:sample} 

A large number of (totaling 112, 191) LRGs were selected from the
Sloan Digital Sky Survey III data release 8 (DR8) based on color and
magnitude criteria \citep[see][]{Eis01}.  By cross-matching the LRG
sample with the cluster catalog given by \citet{Wen12}, we find that
28,336 LRGs are BCGs\footnote{Some BCGs may not locate at the centers
of their host dark matter halos \citep{Skibba11}.}. We also select
those LRGs that are located within a projected separation of $r_{200}$
from the BCGs and with a redshift difference of no more than $0.05$ from the BCGs
\footnote{According to \citet{Wen12}, the photo-z gap of z=0.04(1+z) is
adopted to select member galaxies. Since the redshifts of our sample
galaxies is in the range of 0.15 - 0.25, a redshift gap of about 0.05 
is applied to select the MGs. However, a redshift difference of 0.04
(1+z) is rather large and would introduce some interloper galaxies to
the selected MG sample. To check the effect of this 
contamination, we adopt a more strict selection criterion for 
MGs, i.e.,  the spectroscopic redshift difference between an MG and its
CG is smaller than 0.01, corresponding to a velocity 
difference of $ < 3000 \kms$, which is roughly on the same order 
of the velocity dispersion of massive galaxy cluster.
By this criterion, we re-select a new MG sample, 
which is taken as the most conservative case. We analyze this new MG 
sample and find few differences in the results between those obtained
from the conservative MG sample and the main MG sample.}.
 Here $r_{200}$ is the radius within which the mean
density of a cluster is $200$ times that of the critical density of the
universe \citep[see Table~1]{Wen12}, the redshift difference $0.05$
is adopted to consider the uncertainties in the photometric redshift
estimates; these LRGs are denoted as the `member galaxies' (MGs) of
clusters.  According to the above selection criteria, we obtain 11,703
MGs. The remaining LRGs (72,152), located in environments substantially
less dense than those of the clusters, we denote  as  `field
galaxies' (FGs). To obtain a passively evolving LRG sample, we
 use the following criteria (similar to those in
\citealt{Liu12}):%
\begin{itemize} 

\item having the \texttt{GALAXY\_RED} flag in Catalog Archive Server
(CAS) database;

\item \texttt{zWarning EQ  $0$} , \texttt{$SN_{\rm median} >10$},
\texttt{$0.15 < z < 0.25$},  \texttt{$ph.fracDev\_r > 0.8$} and
\texttt{$V\_disp > 210$},

\end{itemize}
to re-select quiescent LRGs (including BCGs, MGs, and FGs). Here
$SN_{\rm median}$ is the median of the signal to noise ratio (S/N) per
pixel of the whole spectrum. We set zWarning EQ 0 to make sure that
the spectroscopic redshift of each sample galaxy is correct, and the
following analysis  adopts this spectroscopic redshift for each
selected LRG.  Both $SN_{\rm median}$ and $V\_disp$  are taken from
galSpecInfo provided by the CAS database of SDSS and obtained by MPA-JHU
spectroscopic re-analysis. We finally obtain  2,718 BCGs, 1,075 MGs,
and 4,089 FGs. We adopt a small redshift range of $0.15 < z < 0.25$
here  to avoid  the evolution effect. Those LRGs with
obviously $H_\alpha$ and [OII] emission 
(i.e., their $H_\alpha$ and [OII] line emission are larger than zero
at the 2$\sigma$ or above level),
and thus a recent star formation, are removed from the sample 
to obtain a clean sample of passively evolving galaxies
\footnote{Some passively evolving red galaxies may have substantial amounts of
ionized gas and display low-ionization nuclear emission-line region (LINER)-like emission lines. These LINER-like
galaxies ($\sim 19\%$) are also excluded from our sample.}. We
also check the SFHs of those LRGs with H$_{\alpha}$ and [O {\small
II}] emission by using the full spectrum fitting technique, and find
that they have a slightly higher star formation rate at later time
compared with  quiescent LRGs.

\begin{table*}[htbp]
%\begin{table}[htbp] 
%
\centering 
\caption{\label{tab:t1}Quiescent LRGs in different subgroups and
their age properties}
\begin{tabular}{ccccccc} 
\hline \hline \multicolumn{7}{c}{Group I} \\
\hline \multirow{2}{*}{Name}  &  Richness  & Cluster mass   &  $\left<t_{\rm
age}\right>_{\rm m}$ & $\left<t_{\rm age}\right>_{\rm l}$ &
\multirow{2}{*}{Number} &   \multirow{2}{*}{ID} \\ &     $\RL$  & $10^{14} M_\sun$ & Gyr
&  Gyr &      &  \\ \hline 
\multirow{3}{*}{BCGs} & $ 12-22$  & 0.6-1.2 & 9.27$\pm$0.05 & 5.56$\pm$0.05   &
1596 &     1  \\ 
& $ 23-39$ & 1.2-2.4 & 9.39$\pm$0.07 & 5.67$\pm$0.07   & 782   &     2  \\
 & $\geq40$ & $\geq2.4$ & 9.49$\pm$0.11 & 5.85$\pm$0.11  & 340  &      3  \\
\hline \hline
\multicolumn{7}{c}{Group II} \\ \hline 
\multirow{2}{*}{Name}  & $\sigma_v$  & \multirow{2}{*}{$\log M/M_\sun$} & $\left<t_{\rm age}\right>_{\rm
m}$ & $\left<t_{\rm age}\right>_{\rm l}$ & \multirow{2}{*}{Number} &
\multirow{2}{*}{ID} \\ 
&  ($\kms$) &  &  Gyr   &  Gyr &             &  \\ \hline
\multirow{4}{*}{BCGs} & $210-240$  & 11.1-11.4 & 9.09$\pm$0.09 & 5.53$\pm$0.09 &509
&  4 \\ & $240-270$ & 11.4-11.6 & 9.28$\pm$0.06 &5.58$\pm$0.07& 1011 &  5 \\
 & $270-300$ & 11.6-11.8 & 9.34$\pm$0.07 &5.51$\pm$0.07 & 788 &  6 \\
 &  $\geq300$ & $\geq11.8$ & 9.65$\pm$0.10 &5.92$\pm$0.11 &410  &  7 \\ \hline 
\multirow{4}{*}{MGs} & $210-240$  & 11.1-11.4 & 8.90$\pm$0.10 & 5.65$\pm$0.13 & 288
&  8 \\ 
& $240-270$ & 11.4-11.6 & 9.20$\pm$0.10 &5.55$\pm$0.10 &409 &  9 \\
 & $270-300$ & 11.6-11.8 & 9.51$\pm$0.11 &5.89$\pm$0.13&  264 &  10 \\
 &  $\geq300$ & $\geq11.8$ & 9.81$\pm$0.16 &5.93$\pm$0.19  & 114  &  11 \\ \hline
\multirow{4}{*}{FGs}  & $210-240$  & 11.1-11.4 & 8.77$\pm$0.06 & 5.36$\pm$0.06 &
1048 &  12 \\ 
& $240-270$  & 11.4-11.6 & 8.94$\pm$0.05 &5.30$\pm$0.05 &1711 &  13 \\ 
& $270-300$ & 11.6-11.8 & 9.20$\pm$0.06 &5.51$\pm$0.06 &952 &  14 \\
 & $\geq300$  & $\geq11.8$ & 9.40$\pm$0.10 &5.63$\pm$0.10 &  378  &  15 \\ \hline
\hline
\multicolumn{7}{c}{Group III} \\ \hline
\multirow{2}{*}{Name} & $\sigma_v$ & \multirow{2}{*}{$\log M/M_\sun$} & $\left<t_{\rm age}\right>_{\rm
m}$ & $\left<t_{\rm age}\right>_{\rm l}$ & \multirow{2}{*}{Number}  &
\multirow{2}{*}{ID} \\
           &  ($\kms$) &  &  Gyr   &  Gyr &             &  \\ \hline
SCGs  & $>160$ & $>10.6$ & 9.31$\pm$0.08 & 5.60$\pm$0.09   & 508 &  16 \\
 BCGs    &  $>210$ & $>11.1$ & 9.25$\pm$0.04 & 5.64$\pm$0.04  & 2718 &  17 \\
\hline
\end{tabular}
\end{table*}

The above LRG sample is further categorized into several different
groups since the main purpose of this study is to investigate the mass
and environment dependence of the formation of passively evolving
LRGs. 

\begin{itemize}

\item Group I, BCGs are categorized into three subclasses according
to the richness of their host clusters ($R_{\rm L}$).  The richness
boundaries for those subclasses of BCGs are $12 \leq R_{\rm L} \leq
22$, $22<R_{\rm L}\leq 39$ and $R_{\rm L} >39$, respectively, where
$R_{\rm L} =12$ is the low limit for a galaxy system that is classified
as a cluster in \citet{Wen12}. The richness of a cluster can be taken
as a proxy of the cluster mass as $\log M_{200} = (-1.49\pm 0.05) +
(1.17\pm 0.03) \log R_{\rm L}$\citep{Wen12}, where $M_{200}$, in unit of $10^{14}\msun$, is the
cluster mass enclosed in a region with an over-density of $200$.
According to this relation, the above richness boundaries correspond
to the mass ranges of $0.6$-$1.2\times 10^{14}\msun$, $1.2$-$2.4\times
10^{14}\msun$, and $>2.4\times 10^{14}\msun$, respectively. 
 
\item Group II, LRGs with similar masses are categorized into BCGs,
MGs, and FGs, which represent different environments. The mass of an
LRG may be also indicated by its velocity dispersion ($\sigma_v$).
Therefore, BCGs, MGs, and FGs are further categorized into four
sub-groups according to their velocity dispersions. The boundaries for
the velocity dispersion bins are $210$-$240\kms$, $240$-$270\kms$,
$270$-$300\kms$, and $>300\kms$, respectively.  
According to the scaling relation between the stellar mass and the
velocity dispersion of galaxies, e.g., $ \log(M/M_\sun) \approx  0.63
+ 4.52 \log\sigma_v$ \citep{Tho05}, the boundaries of the
$\sigma_v$-bins correspond to the mass ranges of
$10^{11.1}$-$10^{11.4}\msun$, $10^{11.4}$-$10^{11.6}\msun$,
$10^{11.6}$-$10^{11.8}\msun$, and $>10^{11.8}\msun$, respectively.

\item Group III, the satellite galaxies of BCGs (SCGs) are selected
and categorized as a single group, of which the velocity dispersion
range is different from that of the BCGs (see Table~\ref{tab:t1}).

\end{itemize}

We allocate each subgroup with an ID as shown in Table~\ref{tab:t1}.
In the following text, we use these IDs to represent each subgroup, if
not otherwise stated. The number of LRGs in each subclass is listed
in Table~\ref{tab:t1}. 

\section{Spectral synthesis and spectrum fitting}
\label{sec:spectralfit}

\begin{figure*}
%\begin{figure}
\centering
\includegraphics[scale=0.8]{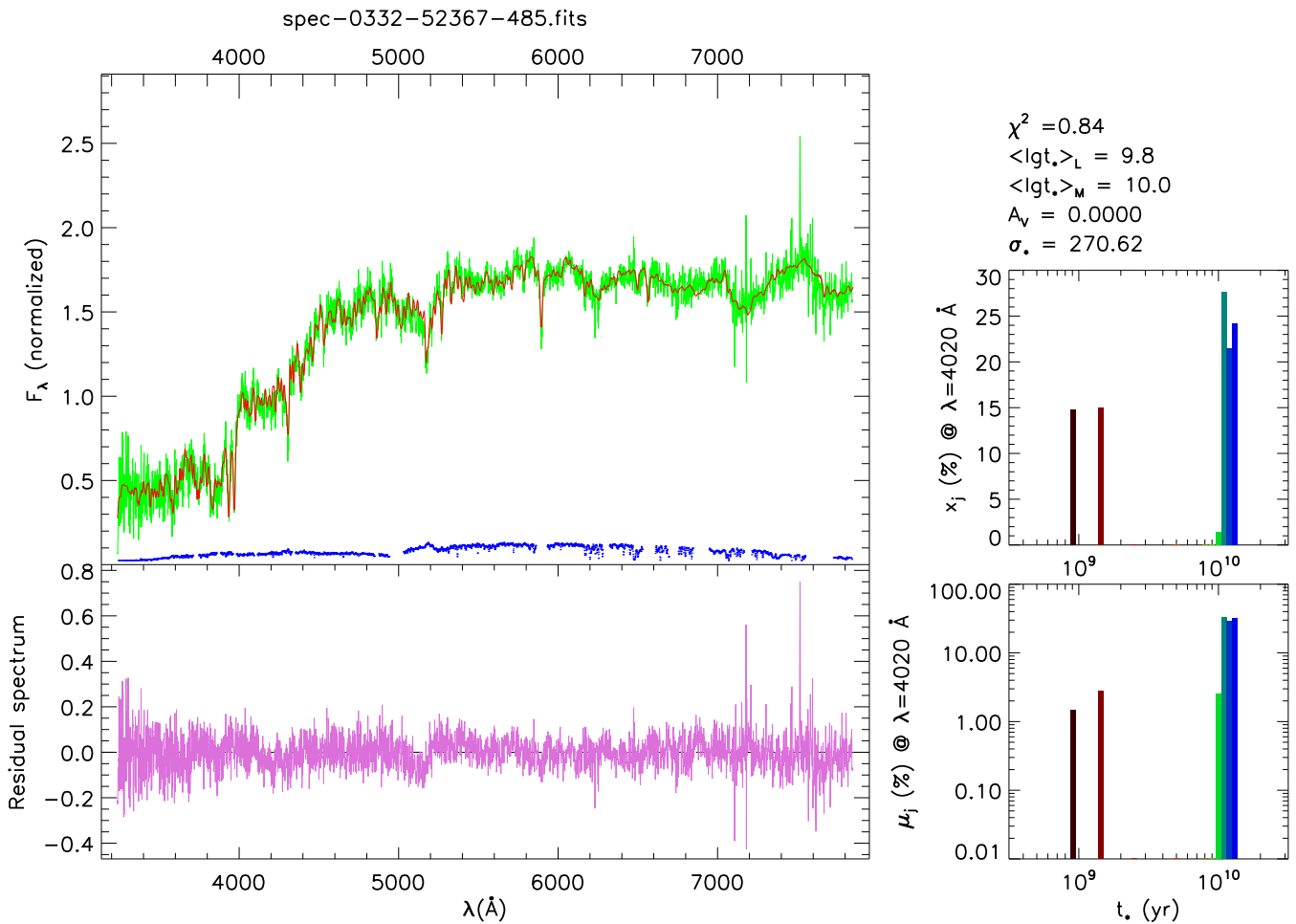}
\caption{Example for the full spectrum fitting of the spectrum of a
BCG.
The green, blue, and red lines
in the top left panel represent the observed spectrum ($O_\lambda$),
its error spectrum, and the model spectrum ($M_\lambda$),
respectively. The residual spectrum ($E_\lambda$) is shown by the
purple line in the bottom left panel. In the left panels, both
$O_{\lambda}$ and $M_{\lambda}$ are scaled by normalizing the flux at
wavelength $4020$\AA\, to 1. The light-weighted and the mass-weighted
fractions of stellar populations, i.e., $x_{\rm j}$ and $\mu_{\rm j}$,
are shown in the top and the bottom right panels, respectively.
Several quantities (see Section~\ref{sec:spectralfit}) derived from
the fitting are listed in the top right corner.  
}
\label{fig:f1}
%\end{figure}
\end{figure*}

In this section, we use the software \texttt{STARLIGHT}
\citep[see][]{Cid05} to derive the stellar populations in each
individual LRG in our sample through  full spectrum fitting. Using
\texttt{STARLIGHT}, the observed SDSS spectrum ($O_\lambda$) of each
sample LRG can be fitted by a model spectrum ($M_\lambda$), a
combination of a set of predefined base spectra, i.e., 
\begin{equation}
M_\lambda = M_\lambda(\vec{x}, A_V, v_\star, \sigma_\star) =
\sum_{j=1}^{N_\star} x_j \gamma_{j,\lambda} r_\lambda,
\label{Mlambda}
\end{equation}
where $\gamma_{j,\lambda} \equiv b_{\lambda,j} \otimes G(v_\star,
\sigma_\star)$, $b_{\lambda,j} \equiv B_{\lambda,j}/B_{\lambda_{0},j}$
is the normalized flux of the $j^{th}$ component, $B_{\lambda,j}$ is
the $j^{th}$ component of the base spectra, $B_{\lambda_{0},j}$ is the
flux of the $j^{th}$ component at wavelength $\lambda_0$, $G(v_\star,
\sigma_\star)$ represents a Gaussian distribution centered at a
velocity $v_\star$ and  broadened by a line-of-sight velocity
dispersion $\sigma_\star$, $x_j$ is the fraction of flux caused by
component {\it j} at $\lambda_0$, and $r_\lambda \equiv
10^{-0.4(A_\lambda-A_V)}$ is the global extinction term represented by
$A_V$. The best-fit is obtained by using a simulated annealing plus
Metropolis-Hastings scheme, which searches for the minimum of
\begin{equation}
\chi^2 = \sum_{\lambda} \left[(O_\lambda - M_\lambda)
w_\lambda\right]^2,
\end{equation}
where $w_\lambda^{-1}$ is the error in $O_\lambda$. The residual
spectrum $E_\lambda$ can also be obtained by subtracting the model
spectrum from the observed one, i.e., $E_\lambda = O_\lambda -
M_\lambda$. 

We first generate the base spectra by taking simple stellar
populations (SSPs) from the Bruzal \& Charlot (2003), hereafter referred to as BC03, evolutionary synthesis models (for
more details, see \citealt{Bru03}). We adopt the Padova 1994
evolutionary tracks \citep{Ber94} and the Salpeter initial mass
function \citep[][]{Sal55}. We adopt ten different SSP ages (i.e.,
0.9, 1.4, 2.5, 5, 8, 9, 10, 11, 12, 13\,Gyr) and four different
metallicity values (0.2, 0.4, 1, and 2.5\,$Z_\odot$), and obtain
a total of 40 SSPs to generate the base spectra. In addition, we  model the
extinction by using the dust extinction law given by \citet{Cal00} 
and the global extinction term $A_{\rm v}$ is a free parameter to be
constrained by the fitting procedure.
We use these 40 SSPs to fit the spectrum of each galaxy in our sample
and extract the information on those SSPs in the formation and
evolution of the galaxy. 

Figure~\ref{fig:f1} shows an example of the fit to the spectrum of a
typical BCG.  The top left-hand panel of Fig.~\ref{fig:f1} shows both the
observed SDSS spectrum $O_\lambda$ (green) and the model spectrum
$M_\lambda$ (red); the bottom left-hand panel gives the residual spectrum
$E_\lambda$ (purple); the top right-hand panel shows the light-weighted
stellar population fractions $x_{\rm j}$; and the bottom right-hand panel
shows the mass-weighted stellar population fractions $\mu_{j}$. As
seen from Fig.~\ref{fig:f1}, the spectrum of this BCG can be well
fitted by a few SSPs with different ages, of which the dominant one in
mass (and light) is an old population with age $\sim 10$~Gyr, and a
young population with age $\sim 1$\,Gyr contributes significantly to
the total light at optical band ($\sim 20\%$), but is almost negligible to
the total mass (less than a few percent).

The fraction of each stellar component, either in mass or in light,
can be obtained through the full spectrum fitting for each galaxy.
Therefore, the age properties and the mean SFHs for each subgroup of
galaxies listed in Table~\ref{tab:t1} can be obtained.  Following
\citet{Cid05}, we define the light-weighted mean age of a galaxy as
\begin{equation}
\langle \log t_\star \rangle_L = \sum _{j=1}^{N_\star} x_j \log t_j,
\label{eq:logtl}
\end{equation}
and the mass-weighted mean age of a galaxy as
\begin{equation}
\langle \log t_\star \rangle_M = \sum _{j=1}^{N_\star} \mu_j \,\log
t_j,
\label{eq:logtm}
\end{equation}
respectively, where $x_j$ and $\mu_j$ are the fractions of flux and
mass contributed by the $j^{th}$ component of the SSPs, respectively.

For the purposes of estimating the Hubble constant by using the
differential age method, however, the mean age should be redefined
later as 
\begin{equation}
\langle  t_\star \rangle_M = \sum _{j=1}^{N_\star} \mu_j \, t_j.
\label{eq:logtm2}
\end{equation}

In principle, the STARLIGHT code can  deal with templates
older than the age of the Universe properly, but the errors in the measurements
($~\sim$ 0.1 dex, caused by the intrinsic degeneracies of stellar populations and measurement uncertainties) may cause  the fitting ages of some galaxies to be larger than the age of Universe ($13.7$\,Gyr), which will be discussed in Section 5 in detail.

\section{Results}\label{sec:results} 

Using the SSP information obtained above for each galaxy, we 
 investigate  statistically the formation and evolution histories of the
LRGs in the different groups listed in Table~\ref{tab:t1}. 

\subsection{Group I: BCGs with different $\RL$ }
\label{sec:comparison1} 

\subsubsection{Age}\label{sec:age1}

\begin{figure} 
\centering 
\includegraphics[scale=0.35]{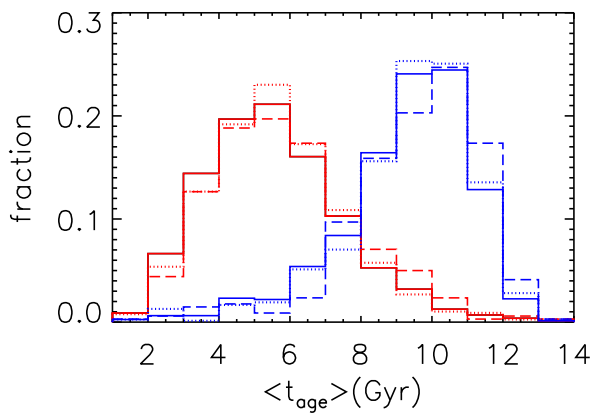}
\caption{Distributions of the mass-weighted (blue histograms) and the
light-weighted ages (red histograms) of BCGs. Histograms shown by the
solid dotted and long dashed lines represent the distributions of
BCGs in galaxy clusters with richness $\RL$ in the range of $12$-$22$,
$22$-$39$, and $>39$, respectively.  }
\label{fig:f2} 
\end{figure}

Figure~\ref{fig:f2} shows the light-weighted (red histograms) and
mass-weighted (blue histograms) mean age distributions of the BCGs in
 subgroups 1, 2, and 3, respectively. The mean ages of these BCGs, which
 are hosted in clusters with different $\RL$, are also listed in
Table~\ref{tab:t1}. According to Fig.~\ref{fig:f2} and
Table~\ref{tab:t1}, the mean mass- or light-weighted age of the BCGs, which are hosted in more massive clusters with larger $\RL,$ appear only slightly
larger than that of those hosted in less massive clusters with smaller
$\RL$, and the differences among the mass-weighted (or light-weighted)
age distributions are statistically insignificant, as indicated by the
Kolmogorov-Smirnov (K-S) tests.  It is apparent that the distributions
of the light-weighted age are substantially offset from those of the
mass-weighted age, mainly because stars with an age less than several Gyr
contribute significantly to the total optical emission but contribute only a
small fraction to the total stellar mass. However, the differences
among the mass-weighted age distributions or  among the light-weighted
age distributions are all insignificant. For simplicity, hereafter we
only show the relevant results for the mass-weighted age, if not
otherwise stated. 

%\begin{figure*}
%\end{figure*}

\subsubsection{Star formation histories}\label{sec:SFH1}
\begin{figure}
\centering
\includegraphics[scale=0.35]{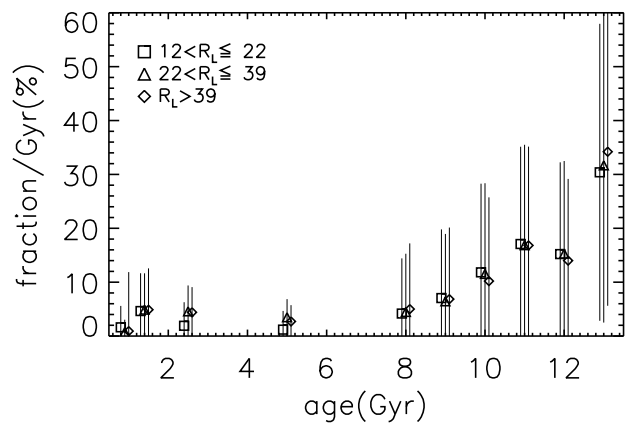}
\caption{Mean mass fractions of SSPs in BCGs.  Squares, triangles, and
diamonds represent the subsamples with $12< \RL \leq 22$, $22 <\RL
\leq 39$, and $\RL>39$, respectively. The bars associated with each
symbol represent the standard deviation of the mean mass fraction. For
clarity, symbols representing an SSP with the same age but for BCGs in
subsamples with $\RL$ from low to high are offset by $\delta
t=-0.1$, $0$, and $0.1$, respectively.  
}
\label{fig:f3} 
\end{figure}

The SSPs with different ages involved in the spectrum fitting of a
galaxy roughly represent the SFH of the galaxy. For each (sub)group
of galaxies, their mean SFH can be obtained by adding together the SSPs in
each galaxy at each age, and as our results show, the
difference in the SSP metallicities (mostly metal rich with $1$ or
$2.5 Z_{\odot}$) is negligible.  Figure~\ref{fig:f3} shows  the mean
mass fractions and the corresponding standard deviations of the SSPs
with each given age, which represent the mean SFH of each sub-group.
As seen from Fig.~\ref{fig:f3},  the mean SFHs of BCGs with
different richness $\RL$ are similar; the old SSPs (with age $\ga
10$~Gyr) are the dominant components to the mean SFHs of the BCGs in
each subgroup; the evolutionary trend of the mean SFHs suggests that
majority of the BCGs obtain most of their masses in a relatively short
period at  early time $\ga 10$\,Gyr; several young SSPs with age $\la
1$\,Gyr do contribute to these BCGs, although they appear to be
insignificant in the mass fraction ($\la $ a few percent); and the
later star formation appears to be extended and more or less at a
constant rate over a period of more than several Gyr.

%\begin{figure*}
%\end{figure*}

\subsubsection{Young, intermediate and old SSPs}\label{sec:SSPs1}

\begin{figure*}
%\begin{figure}
\centering
\includegraphics[scale=0.7]{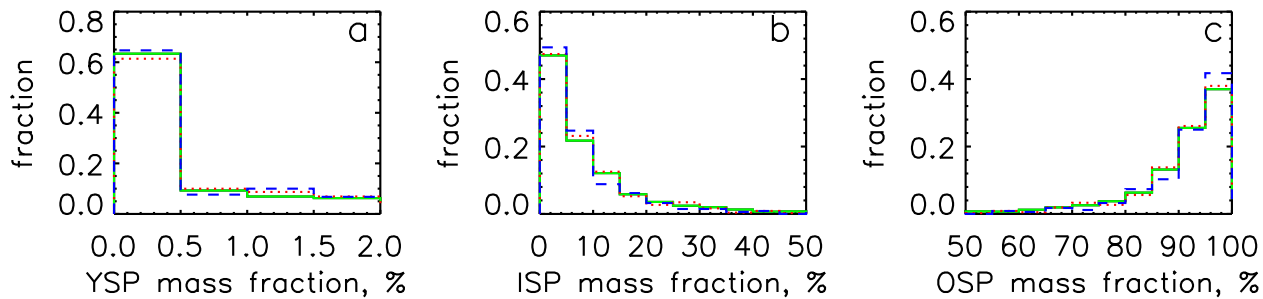}
\caption{Distributions of the mass fraction of the YSP, ISP, and OSP
in BCGs.  Panels (a), (b), and (c) show the distributions of the mass
fraction of the YSP, ISP, and OSP, respectively.  In each panel,
histograms shown by the (green) solid, (red) dotted, and (blue) dashed
lines represent the distributions of the sub-groups with $12 < \RL
\leq 22$, $22<\RL \leq 39$, and $\RL>39$, respectively. }
\label{fig:f4} 
%\end{figure}
\end{figure*}

We divide the SSPs into three groups according to their ages, i.e.,
(1) the young SSP (YSP), including the SSPs with an age $\leq 1$\,Gyr;
(2) the old SSP (OSP), including those SSPs with an age $>2.5$\,Gyr; and
(3) the intermediate SSP (ISP), including those SSPs with an age in
between the YSP and the OSP.  Figure~\ref{fig:f4} shows the histograms
for the distributions of mass fractions of YSP, ISP, and OSP in the
BCGs with different richness, respectively.  As shown in
Fig.~\ref{fig:f4},  the mass fractions of the YSP in most BCGs are
small ($\leq 0.5\%$), while the mass fractions of the OSP in most BCGs
are large ($\geq90\%$). There are no apparent differences among the
distributions of the mass fraction of the YSP, or the ISP, or the OSP,
in the BCGs with different richness. 

According to the analysis in this section, we conclude that the masses
of the host clusters of BCGs have little effect on the SFHs and the age
properties of those BCGs.

\subsection{Group II: BCGs, member and field galaxies}
\label{sec:comparison2}

To  consider the mass and environmental effects on the
formation of LRGs further, we explore the SFHs of the LRGs
that are categorized as BCGs, MGs, and FGs (see Group II in
Table~\ref{tab:t1}). Galaxies in different $\sigma_v$ bins are in
different mass ranges as the mass of a galaxy is roughly proportional
to the fourth power of $\sigma_v$ \citep[e.g.,][]{FJ76}, and the
masses of those BCGs, MGs, and FGs are approximately the same in a
given $\sigma_v$ bin.  However, the LRGs in the largest $\sigma_v$
bin ($\sigma_v > 300\kms$) are roughly $6  $ or more times  massive than
those in the smallest $\sigma_v$ bin ($210\kms <\sigma_v < 240\kms$). 

\begin{figure}
\centering
\includegraphics[scale=0.35]{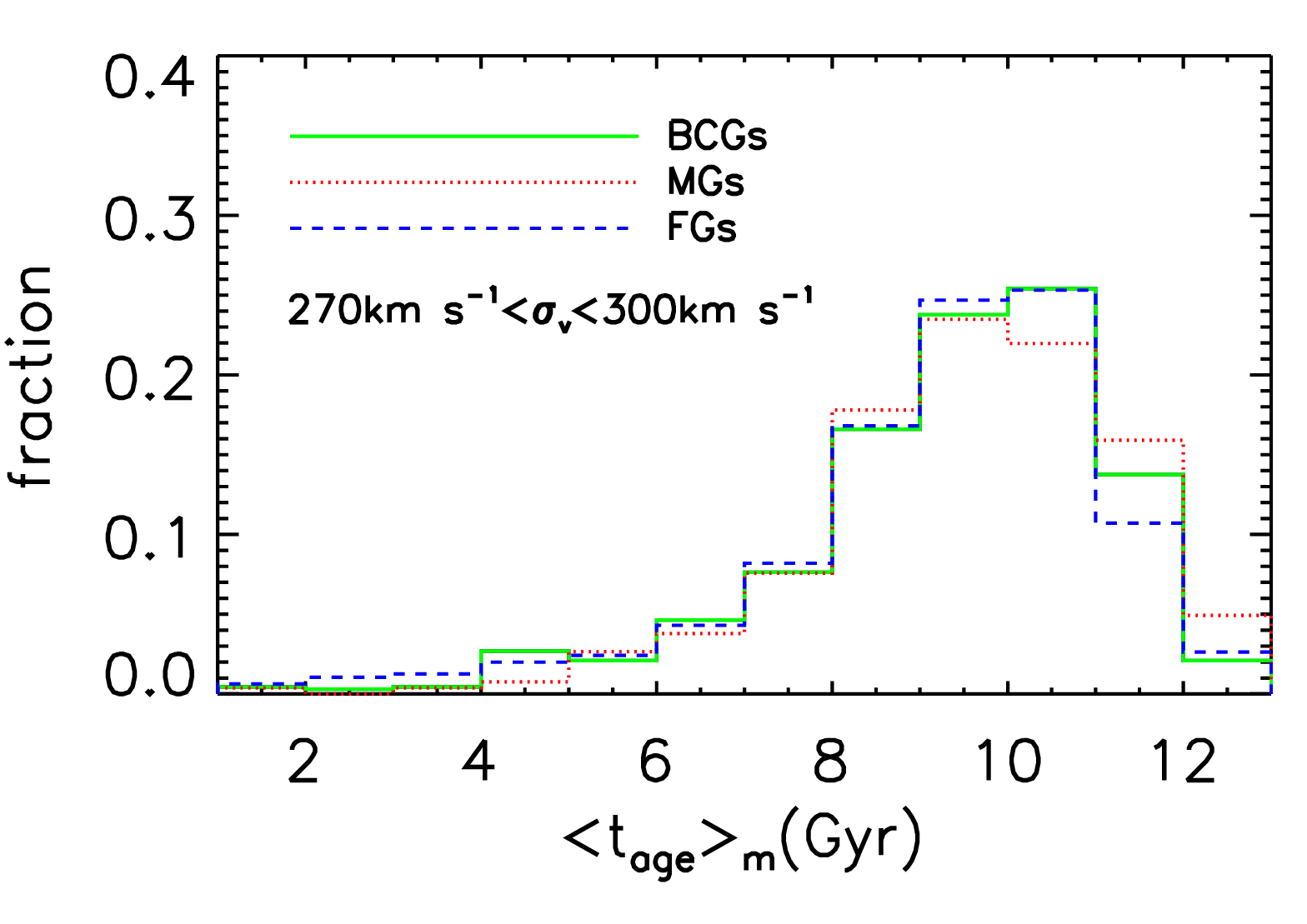}
\caption{Age distributions of BCGs (green solid line), MGs (red dotted
line), and FGs (blue dashed line). The velocity dispersions of those
BCGs, MGs, and FGs all span  the range from $270\kms$ to $300\kms$.
}
\label{fig:f5} 
\end{figure}

Figure~\ref{fig:f5} shows the age distributions of BCGs (green solid
line), MGs (red dotted line), and FGs (blue dashed line), with
$\sigma_v$ in the range from $270\kms$ to $300\kms$, respectively.
Apparently, there are no significant differences among the age
distributions of these three subgroups, although BCGs and MGs appear
to be slightly older than FGs. The mean mass-weighted ages of BCGs,
MGs, and FGs are $9.34\pm 0.07$, $9.51\pm 0.11$, and $9.20\pm
0.06$\,Gyr, respectively. For each of the other $\sigma_v$ bins, 
as shown in Appendix A , the differences among the age distributions
of the BCGs, MGs, and FGs are also not significant, similar to those
shown in Fig.~\ref{fig:f5} for the bin of $270\kms< \sigma_v <
300\kms$. Part of reason that BCGs have a mean age similar to that of
MGs might be that some BCGs are actually satellite galaxies rather
than the central galaxies, which causes a mixing of the central
galaxies with satellite galaxies in BCGs \citep{Skibba11}.

\begin{figure}
\centering
\includegraphics[scale=0.35]{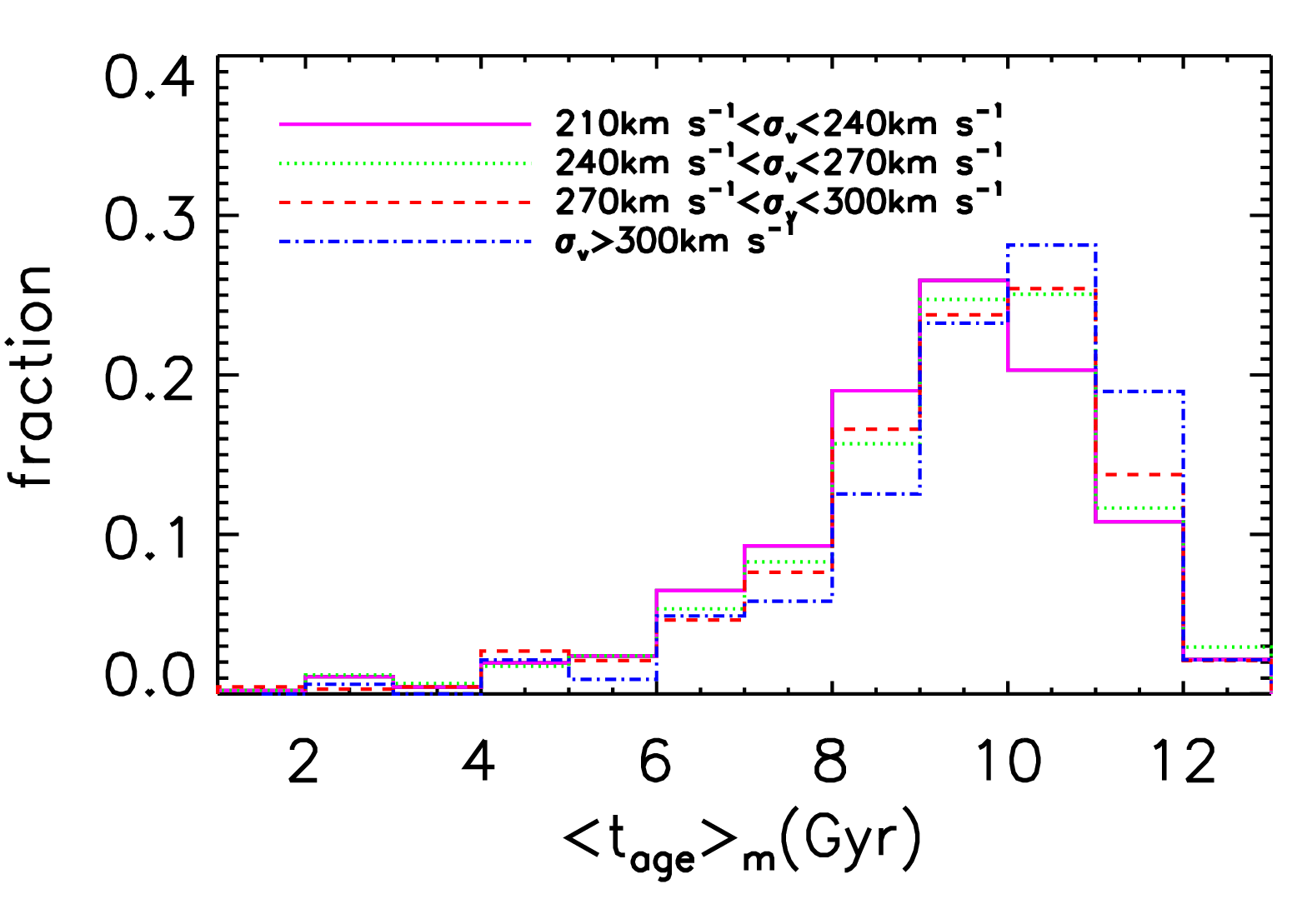}
\caption{Age distributions of BCGs in different velocity dispersion
ranges. Histograms shown by the solid (magenta), dotted (green),
dashed (red), and dash-dotted (blue) lines represent the age
distribution for BCGs with velocity dispersions in the range of
$210\kms<\sigma_{v}\leq240\kms$, $240\kms<\sigma_{v}\leq270\kms$,
$270\kms<\sigma_{v}\leq300\kms$, and $\sigma_{v}\geq300\kms$,
respectively. }
\label{fig:f6} 
\end{figure}

Figure~\ref{fig:f6}, as an example, shows the age distribution of
BCGs in different $\sigma_v$ bins. There is a weak tendency that the
ages of the BCGs in a larger $\sigma_v$ bin are slightly older than
those of the BCGs in a smaller $\sigma_v$ bin, though the difference
is small. The mean age of the BCGs in the velocity dispersion bins of
$210$-$240$, $240$-$270$, $270$-$300$, and $>300\kms$ are $9.09\pm
0.09$, $9.28\pm 0.06$, $9.34\pm 0.07,$ and $9.65\pm 0.1$\,Gyr,
respectively (see Table~\ref{tab:t1}). The apparent three-sigma difference of the means
are due to  the statistical errors that we adopt here are the errors for the
means (i.e., $\sqrt{\sum (age -mean)^2/(N*(N-1))}$) rather than the scatters
(i.e.,$ \sqrt{\sum (age-mean)^2/N}$) of the distributions. For MGs and FGs, we also find
the tendency that their ages increase slightly with growing
velocity dispersion, the same applying to BCGs (see also
Table~\ref{tab:t1}).  We also perform the K-S test for the age
distributions of BCGs within each of the four different velocity
dispersion bins (subsamples~4, 5, 6, and 7), and find that the
probability that any two of them draw from the same distribution is
larger than $99\%$.

%\begin{figure*}

\subsubsection{Star formation histories}\label{sec:SFH2}

\begin{figure*}
%\begin{figure}
\centering
\includegraphics[scale=0.7]{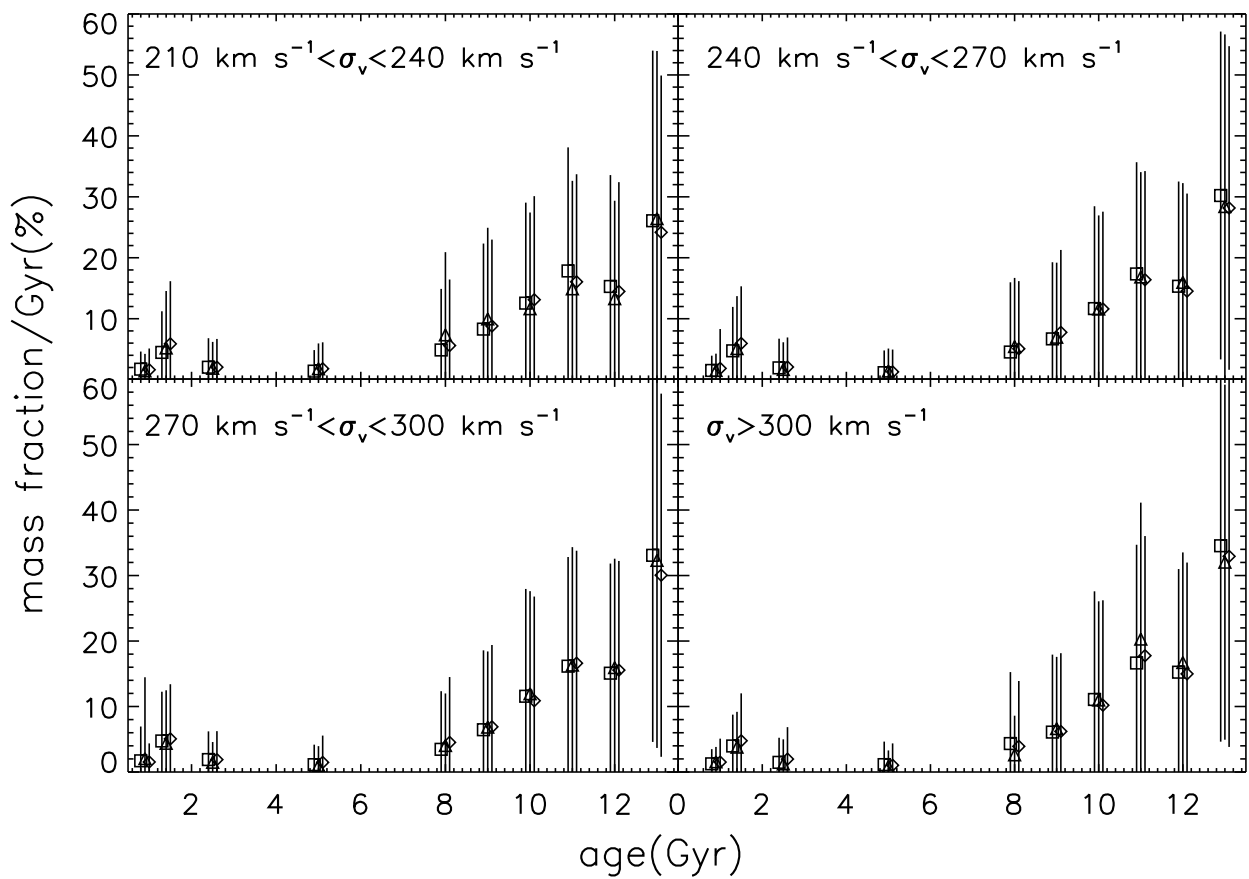}
\caption{Mean mass fraction of SSPs with different ages for the
subsamples with different velocity dispersions. Squares, triangles,
and diamonds represent the mean mass fraction of SSPs for BCGs, MGs, and FGs, respectively.
The bar associated with each symbol represents the standard deviation
of the mean mass fraction. For clarity, symbols representing an SSP
with the same age but for different sub-samples are offset by $\delta
t=-0.1$, $0$, and $0.1$, respectively. 
}
\label{fig:f7} 
%\end{figure}
\end{figure*}

We also investigate and compare the SFHs of BCGs, MGs, and FGs in each
$\sigma_v$ bin, according to the SSPs obtained from the full spectrum
fitting for each object. Figure~\ref{fig:f7} shows the mean mass
fractions of SSPs formed per Gyr in the BCGs, MGs, and FGs for the
four different $\sigma_v$ bins, respectively, which represent the mean
SFHs of those galaxies. As seen from Fig.~\ref{fig:f7}, the mean
SFHs of the BCGs, MGs, and FGs in each given $\sigma_v$ bin are almost
the same, and there is little difference between the SFHs of BCGs (or
MGs, or FGs) in different $\sigma_v$ bins, which suggests that the
environmental dependence is weak and the mass dependence of the SFHs
of quiescent LRGs is also not statistically significant.
On average, most of the stars ($\ga 90\%$) in BCGs, MGs, and FGs were
formed about $8$\,Gyr ago; and less than $10\%$ stars formed
after $8$\,Gyr via a more or less constant rate; and the mean fraction
of young SSP with an age $< 1$\,Gyr is usually $<0.5\%$ and can almost be
neglected. 

\subsubsection{Young, intermediate and old SSPs}\label{sec:SSPs2}

\begin{figure*}
%\begin{figure}
\centering
\includegraphics[scale=0.7]{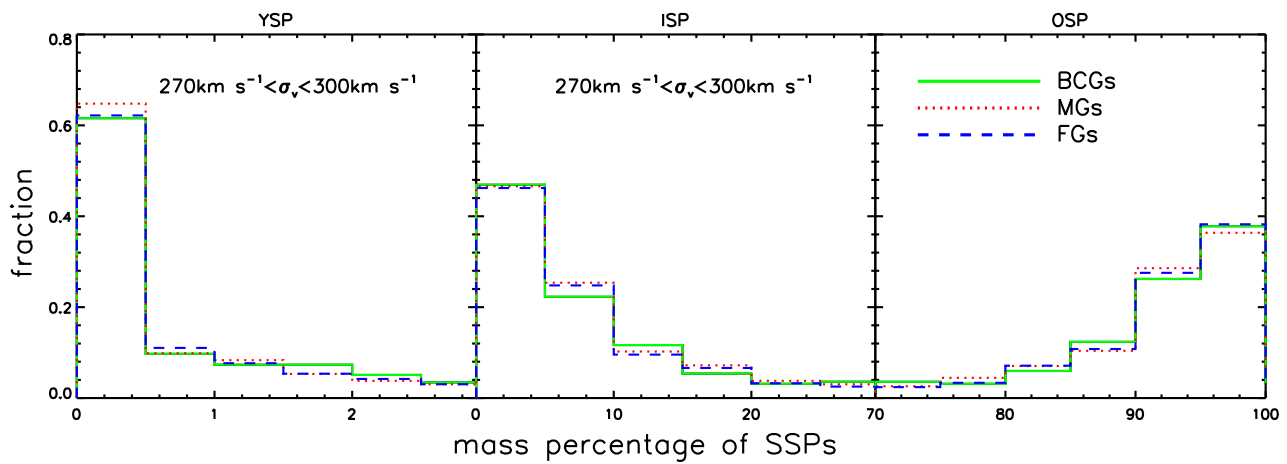}
\caption{Distributions of the mass fraction of different SSPs for LRGs
in different subsamples. As an example, only the distributions
obtained for those galaxies with velocity dispersions $\sigma_v>300
\kms$ are shown here.  Histograms shown by the solid (green), dotted
(red), and dashed (blue) lines represent the distributions for BCGs,
MGs, and FGs, respectively.  }
\label{fig:f8} 
%\end{figure}
\end{figure*}

We compare the YSP, ISP, and OSP mass fractions for BCGs, MGs, and FGs,
similar to the comparison done in Section~\ref{sec:SSPs1}, to
illustrate the detailed distribution of SSPs in each subgroup of
galaxies.  Figure~\ref{fig:f8} shows the distributions of the mass
fractions of YSP (left-hand panel), ISP (middle panel), and OSPs (right
panel) among those BCGs (green line), MGs (red line) and FGs (blue
line) with $\sigma_v$ in the range from $270\kms$ to $300\kms$. (For other $\sigma_v$ bins, the
distributions of YSP, ISP, and OSP are shown in Appendix A. These are
similar to those shown in Fig.~\ref{fig:f8} for the bin with
 $\sigma_v$ in the range from $270\kms$ to $300\kms$.)
 We find that the mass fraction of YSP in majority of BCGs, MGs, or
FGs is $<0.5\%$ (as seen from the left-hand panel of Fig.~\ref{fig:f8}),
the mass fraction of ISP in majority of the BCGs/MGs/FGs is $<10\%$,
and the mass fraction of OSP in majority of the BCGs, MGs, or FGs is
$>90\%$. 

\subsection{Group III: BCGs and their associated MGs}
\label{sec:comparison3}

In Section~\ref{sec:comparison2}, we have explored the SFHs and the
age properties of BCGs, MGs, and FGs, where the MGs are the member
galaxies of the host clusters of the BCGs, and they are categorized
into sub-groups with $\sigma_v$ in the same range as that of the BCGs.
Many MGs with small $\sigma_v$ may exist in the host clusters of a BCG
with large $\sigma_v$; and some MGs in a given $\sigma_v$ bin may
actually be the member galaxies of the host clusters of some BCGs
grouped in a higher $\sigma_v$ bin. We further investigate and compare
the SFHs and age properties of these BCGs and their associated member
galaxies (i.e., the SCGs in Table~\ref{tab:t1}), without considering
the differences of their velocity dispersion.  The velocity
dispersions of SCGs are systematically smaller than those of the BCGs.
By comparison of both the SFHs and the age properties of the BCGs and
the SCGs, as for those  in Section~\ref{sec:comparison2}, we find that
the SFHs and age properties of SCGs are also similar to those of BCGs,
and statistically there is no apparent difference between the SFHs/age
properties of BCGs and those of SCGs.

\subsection{Validity of the population synthesis results}
\label{sec:validation}

The results obtained from the full spectrum fittings may depend on the
initial settings of the base spectra. To close this section, we
 check the validity of the results obtained from the population
synthesis above by  exploring various different settings for the
base spectra further. First, we  alternatively adopt either fourteen different
SSP ages (i.e., $0.9, 1.4, 2, 3, 4, 5, 6, 7, 8, 9, 10, 11, 12,$ and
$13$\,Gyr) or eight different SSP ages (e.g., $0.9, 1.4, 2.5, 3, 5, 8,
10,$ and $13$\,Gyr) and four different metallicity values ($0.2, 0.4,
1,$ and $2.5$\,$Z_\odot$) to generate the base spectra. We redo the
fits for all sample galaxies by using these base spectra. We find that
there are no apparent differences in the resulting age distributions of
SSPs in each velocity dispersion bin.  These tests suggest that our
results are not sensitive to the base spectra.

The results obtained from the population synthesis model may also
depend on the adopted initial mass function (IMF). To check this, we alternatively adopt the
Chabrier IMF and redo the fits for all sample galaxies. We find that
the mean mass fractions of SSPs with different ages for BCGs, MGs, and
FGs in each velocity dispersion bin are not much different from those that result from a Salpeter IMF, although  the  ages of the
LRGs  that are derived from using the Chabrier IMF are slightly older than those resulting from
the Salpeter IMF. We also replace the BC03 model by the CB07 model to
re-fit the LRG spectra and find that the results are almost the same.

The existence of multiple solutions is a well known problem in stellar
population synthesis. It is necessary to assess the reliability of the
model parameters that result from the full spectrum fittings, which may
be quantified by Monte Carlo simulations.  \citet{Cid05} have done
such simulations and demonstrate that Starlight can produce reliable
estimates of $\langle \log t_{\star}\rangle_M$ and $A_v$.  They also
investigate the age--metallicity degeneracy and find that this
degeneracy may induce systematic biases in the estimates of $\langle
Z_\star\rangle$ and $\langle \log t_{\star}\rangle$ at the level of
$\la 0.1-0.2$\,dex.  The statistical differences among the formation
and evolution histories of different groups of LRGs in this paper are
not affected by this level of precision, but the systematic biases
in the age estimates will affect the constraints on the Hubble
constant, as discussed in Section 6.

To close this section, we note  that several previous studies
have investigated the ages of early-type galaxies as a function of their
environment and masses. \citet{Ber06} studied the photometric and
spectral properties of SDSS early-type galaxies as a function of their
local environment and redshift, and they found that galaxies located
in the densest regions are typically 1 Gyr older than those galaxies (with the
same luminosities) in less dense regions. However,
environment is quantified in two different ways in \citet{Ber06},
either by the distance to the nearest cluster of galaxies, or by the
distance to the tenth nearest luminous neighbor, which are different
to those in our paper; and their sample includes galaxies with strong
OII emission, or with $H_\delta$ and $H_\gamma$ Balmer absorption lines (however,
these galaxies are excluded from our sample), which indicates significant star formation in
the not very distant past.  \citet{Tho05} analyze more than 100 early-type galaxies, including a roughly equal number of
elliptical and lenticular (S0) galaxies, and they find that massive
early-type galaxies in low-density environments appear to be ~two Gyr
younger than their counterparts in high-density environments. In our
study, most LRGs should be early-type galaxies but not S0.
\citet{Tho10} study the effect of environment on the evolution of
early-type galaxies by analyzing the stellar population properties of
a large number of early-type massive galaxies selected from SDSS after a
visual inspection of their morphologies, and they find that the
influence of the effect of the environment increases with decreasing galaxy
mass. \citet{Ber98} found that the mass-weighted age differences
between cluster and field elliptical galaxies could be significantly 
smaller than 1 Gyr and brighter field elliptical galaxies appear to 
be similar to their cluster counterparts.
The latter two papers appear roughly consistent with our results because most
LRGs studied in this paper are bright and massive early-type galaxies.

\section{Comparison with the results from the semi-analytical model of
galaxy formation}
\label{sec:compsammodel}

According to the analysis above, there are few differences between
the age distributions and the mean SFHs of quiescent LRGs in different
environments, which suggests that these LRGs, whether located in
dense or less dense environments, can be securely used as cosmic
chronometers.  However, it appears that the little environmental
dependence of the age properties of quiescent LRGs is in contradiction
with basic expectations. For a deeper understanding of this problem,
below we adopt the semi-analytical modeling of galaxy formation by
\citet{Guo11} to select mock LRGs, and compare the age properties and
the SFHs of the mock LRGs with the observational ones obtained
above.

We select LRGs from the mock galaxies that were generated by \citet{Guo11}
using the following criteria, i.e., $M_{\ast}>10^{11} M_{\odot}$,
$M_{\rm bulge}/M_{\ast} > 0.8$ and $|C_{\bot}|<0.2$ at $z\sim0$, where
$|C_{\bot}|$ is defined as $|C_{\bot}|= (r-i) - (g-r) /4-0.18$,
$M_{\rm bulge}$ and $M_{\ast}$ are the mass of the bulge of a galaxy
and the galaxy, respectively, $g$, $r$, and $i$ represent the
$g$-band, $r$-band and $i$-band  magnitudes, respectively. These
criteria were used to select LRGs in SDSS by \citet{Eis01}.

Using a similar approach as for the SDSS LRGs, we also categorize the LRGs
that have been selected from the mock galaxies into three groups, i.e., BCGs, MGs, and
FGs, respectively. BCGs are the central galaxies in massive clusters
that are hosted in dark matter halos with masses $M_{\rm H} \geq 6 \times
10^{13} M_\sun$ (corresponding to the low limit of richness $\RL=12$
for the clusters selected in \citealt{Wen12}); MGs are massive
satellite LRGs that are hosted in the same dark matter halos of BCGs; and FGs
are the biggest galaxies in the dark matter halos with mass in the
range of $10^{13}M_\sun < M_{\rm H} <6\times 10^{13}M_\sun$,
where $10^{13} M_\sun$ is set as the lower limit of the halo mass
since (1) many studies indicate that the typical halo masses of LRGs
are about a few times $10^{13}h^{-1}M_\sun$ or larger (\citealt{Guo14,
Par13,  Hud15}) and (2) \citet{Man06} found that the halos with mass
$\ga 10^{13}M_\sun$ host elliptical galaxies at their center while even the
brightest central spirals are hosted by halos with masses
$<10^{13}M_\sun$.
For each LRG, we trace its merger and assembly history back to high
redshift. 

The SFHs of galaxies are closely related to their assembly histories.
In the semi-analytic models (SAM), most of the large elliptical galaxies, presumably
massive LRGs or BCGs in the centers of the most massive halos, are
formed through (many) major mergers of smaller progenitor galaxies,
and most of the cold gas in each of them is consumed via a starburst
triggered by the merger. 
These galaxies experienced little star formation
after the last (wet) major mergers they experienced. Some satellite
galaxies may transform to ellipticals due to the gravitational
perturbation from their main halos. These satellite galaxies (MGs) may
also have little gas to be cooled down to form new stars, after they
fell into the main progenitor halos, because most of the hot gas in
their progenitor halos may have been stripped off by their main halos.  Less
massive LRGs (or elliptical galaxies; FGs) may be hosted in less
massive halos (e.g., with mass in the range of $10^{13}M_\sun < M_{\rm
H} < 6 \times 10^{13}M_\sun$), they may experience less frequent major
mergers in their assembly histories. Because the assembly histories of the
above three populations of galaxies may be different, therefore the
SFHs of these three populations may be different as well. Below we use the
assembly histories extracted from the SAM model by \citet{Guo11} for
 different populations of mock galaxies, i.e., BCGs, MGs, and
FGs, to check the differences.
\begin{figure}
\includegraphics[scale=0.7]{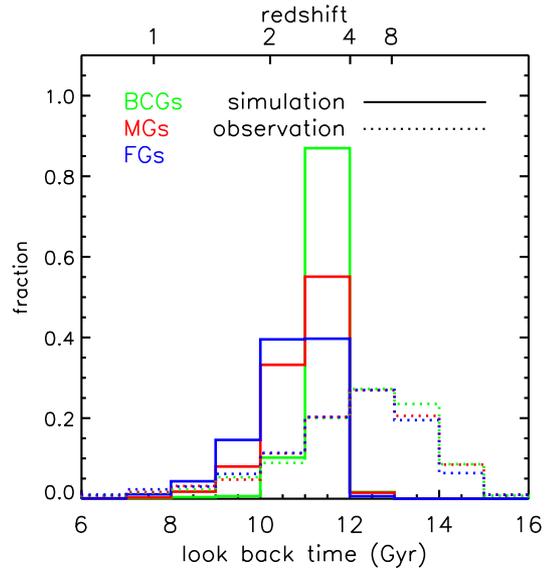}
\caption{Distribution of the mass-weighted ages for sample galaxies
in different types. The green, red, and blue lines represent the
samples of BCGs, MGs, and FGs, respectively.  The thin dotted lines
represents the age distribution obtained for LRGs with velocity
dispersion $>210\kms$
by the full spectrum fitting in this paper, and the thick solid lines
represent the age distributions obtained from a semi-analytic galaxy
formation model \citep{Guo11}. Some SDSS LRGs appear to
have ages or lookback times of 15 Gyr, larger than the Hubble time,
because of  measurement errors in the STARLIGHT-derived ages.} 
\label{fig:f9}
\end{figure}

Figure~\ref{fig:f9} shows the mass-weighted age distributions of
these mock BCGs, MGs, and FGs, respectively. Apparently, the age distribution of
BCGs is limited in a small range from $10$\,Gyr to $12$\,Gyr (or from
redshift $z \sim 2$ to $4$); the age distribution of MGs is more
extended and skewed towards small values of $\sim 8$\,Gyr, compared
with that of BCGs; and the age distribution of FGs is even more
extended and also skewed toward small values of $\sim 8$\,Gyr. The
mean age of the BCGs, MGs and FGs are $11.4\pm 0.006$, $11.0\pm 0.007,$ and
$10.6\pm 0.004$\,Gyr, respectively. The mean age of MGs is slightly
younger than that of BCGs by $0.4$\,Gyr and older than that of FGs by
$0.4$\,Gyr, and these differences are not significant considering the
wide associated scatters.  Figure~\ref{fig:f9} also shows the
mass-weighted age distributions of the corresponding groups of the
SDSS LRGs with $\sigma_v\geq 210\kms$
(histograms in dotted lines). We note that the mock galaxies are selected
at redshift $z=0$, while the SDSS LRGs are selected in the redshift
range from $z=0.15$ to $0.25$.  To account for this difference, we
assume no star formation in those galaxies at $z\la 0.15-0.25$ and
re-calculate the stellar mass-weighted age by replacing
Equation~(\ref{eq:logtm}) with
\begin{equation}
\left< \log t_\star \right> = \sum^{N_\star}_{j=1} \mu_j \log (\delta
t_j + t_j),
\end{equation}
where $\delta t_j$ is the lookback time of the galaxy $j$ at $z_j$.
We adopt a flat $\Lambda$CDM cosmology with $H_0= 73 \kms{\rm
Mpc}^{-1}$ and $\Omega_{\rm m}= 0.25$ as used in \citet{Spr05}.  The
mean ages of the SDSS BCGs, MGs and FGs are $12.1\pm 0.04$, $12.0\pm 0.06$ and
$11.8\pm 0.03$\,Gyr, respectively. It appears that the mean age of MGs
is still slightly younger than that of BCGs, but older than that of
FGs, respectively. The differences between the age distributions of
those observed LRGs are less obvious than the ones seen for the mock LRGs.
However, it is obvious that the age distributions obtained for the
observed BCGs, MGs, and FGs are all broader than the corresponding
distributions obtained from the SAM model, and the mean age of the
observed ones are older than those obtained from the SAM model by
$\sim 1.0$\,Gyr, which might suggest that the mass-weighted ages of
the observed LRGs, which  were estimated  using  STARLIGHT, are systematically
overestimated.

\begin{figure}
\includegraphics[scale=0.7]{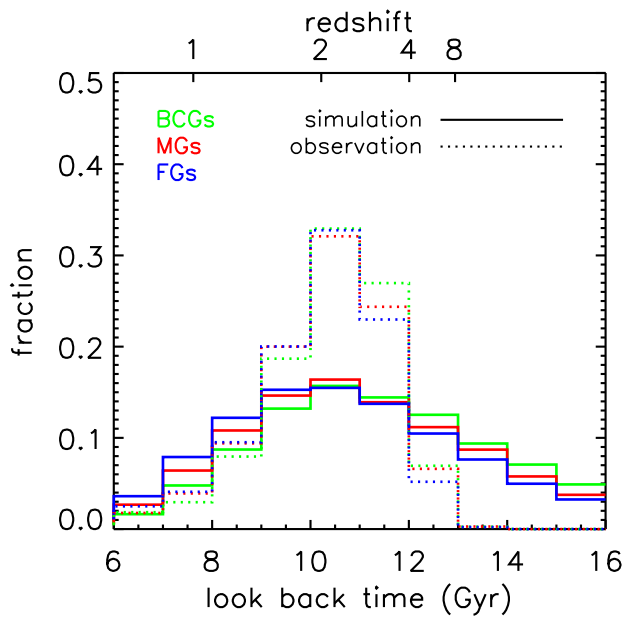}
\caption{Comparison between the mass-weighted ages obtained from
simulation and those obtained from observations by considering the measurement
errors.  The green, red, and blue lines represent the samples of BCGs,
MGs, and FGs, respectively.  The thin dotted lines represent the age
distributions obtained for different LRG samples by the full spectrum
fitting in this paper, after adding a correction of $-0.09$\,dex, and
the thick solid lines represent the age distributions obtained from a
semi-analytic galaxy formation model \citep{Guo11} by considering a
scatter of 0.1 dex.
%after allocating a 0.1 dex normal distribution error. 
}
\label{fig:f10}
\end{figure}

We note that \citet{Gon10}  investigate the accuracy of the age
estimates by using full spectrum fitting. They apply  STARLIGHT
to a sample of globular clusters, fitting the cluster-integrated light
spectra with a suite of modern evolutionary synthesis models, such as
the Galaxev/STELIB,  as used in this study. By comparing the results
obtained from the full spectrum fitting with those of other methods in
the literature, they conclude that the ages estimated by the full
spectrum fitting are older than those estimated  by the S-color magnitude diagram
(S-CMD) method \citep{Girardi95} by $\sim 0.09$\,dex, and the accuracy
of these age estimates is about $\sim 0.1$\,dex. 
Galaxies are more complicated than globular cluster because of
their SFHs and the errors of the STARLIGHT-derived ages for  LRGs may be
different from those for globular clusters.  However, there is no such
study on the possible biases or errors in the STARLIGHT-derived ages
for galaxies or LRGs in the literature. Here we simply assume that the
biases or errors in the STARLIGHT-derived ages for LRGs are the same
as that for globular clusters, and 
we allocate an error following a normal distribution with standard
deviation of $0.1$\,dex to the mass-weighted age of each mock LRG, and
add a correction of $-0.09$\,dex to the age obtained from the full
spectrum fitting for each observed LRG. We reobtain the age
distributions for both the mock and the observed BCGs, MGs, and FGs,
as shown in Fig.~\ref{fig:f10}. By these corrections, we find
that the age distributions for the mock BCGs, MGs, and FGs appear to
 match well those obtained from the observations. 

\begin{figure}
\includegraphics[scale=0.7]{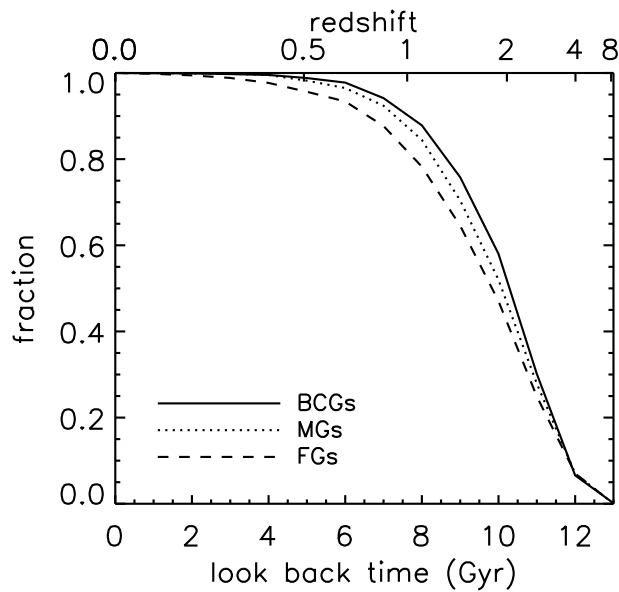}
\caption{Accumulated distribution of the quenching time of the
simulated LRGs. The quenching time of a galaxy here is set as the lookback time of the universe at which the galaxy just
experienced its last wet major merger or became a satellite.  The
solid, dotted, and dashed lines represent the distribution of BCGs,
MGs, and FGs, respectively.  }
\label{fig:f11}
\end{figure}

We further investigate the SFHs and the quench time of the mock BCGs,
MGs, and FGs, to reveal the underlying physics for the small
differences between the distributions of the ages and the SFHs of the
observed BCGs, MGs, and FGs, as illustrated in
Section~\ref{sec:results}. For a BCG or an FG, the time at which the
star formation in it is quenched can be represented by the time of the
last wet major merger it experienced; while for an MG, the quench time
is represented by the earlier one of the last (wet) major merger time
and the infall time. For the mock galaxies that result from the SAM
model, we assume a merger as a wet major merger if the mass ratio of
the two progenitor galaxies is in the range of $1/3$ to $3$ and the
special star formation rates in the two galaxies are both $>0.01{\rm
Gyr}^{-1}$.  Figure~\ref{fig:f11} shows the distribution of the
quench time for BCGs, MGs, and FGs, respectively. As seen from the
figure, LRGs in dense environment, including BCGs and MGs, are
quenched statistically earlier than those FGs in less dense
environments; and FGs appear to be quenched slightly later than BCGs.
Roughly half of BCGs and FGs experienced their last
wet major mergers before $10.3$\,Gyr and $9.6$\,Gyr ago, respectively, and about half of MGs
experienced their last wet major mergers or fell into main halos
over $10$\,Gyr ago.  However, the systematic
differences between the distributions of the quench time of BCGs, MGs,
and FGs
%, offset by $\sim 0.3$ to $0.7$\,Gyr, 
are relatively small compared with the scatters of those found for
the SDSS LRGs as listed in Table~\ref{tab:t1}. 

\begin{figure}
\includegraphics[scale=0.7]{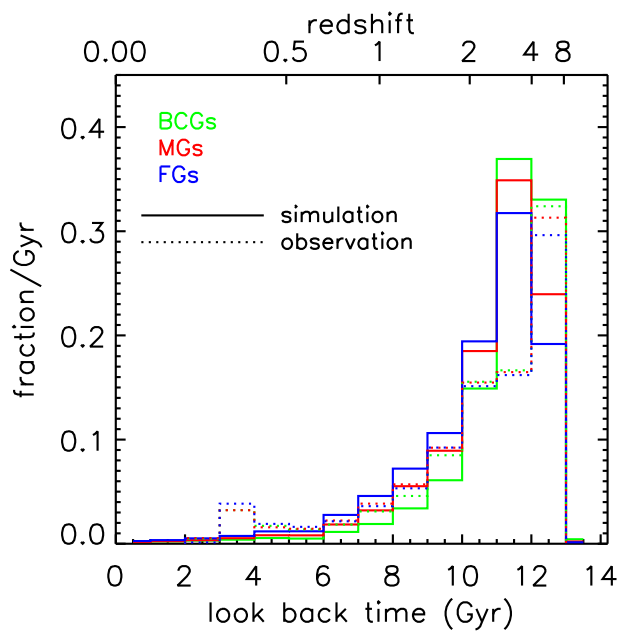}
\caption{Mean star formation histories of different types of LRGs. The
SFH of a galaxy is represented by the mass ratio of stars formed per
Gyr around any given lookback time $t$ to the total stellar mass.  The
solid green, red, and blue histograms represent the mean value of the
mass growth history for BCGs, MGs, and FGs obtained from a
semi-analytical model \citep{Guo11}, respectively; the dotted
histograms are obtained from the full spectrum fitting for the SDSS
BCGs, MGs, and FGs, respectively. }
\label{fig:f12}
\end{figure}

The mean SFH of those galaxies in each subsample can be represented
by the evolution of the mass fraction of the stars formed per Gyr
around a given lookback time, as shown in Figures~\ref{fig:f3} and
\ref{fig:f7}. By recording the mass of formed stars in each mock
galaxy at each snapshot, the SFH of each mock galaxy can be obtained,
and consequently the mean SFH of any group of mock galaxies can also
be obtained. Figure~\ref{fig:f12} shows the mean SFHs obtained from both
 the SAM model (solid lines) and the SDSS observations (dotted
lines) for BCGs, MGs, and FGs, respectively. For the SFHs obtained
from observations through STARLIGHT, a correction of $-0.09$\,dex has
been adopted for the age estimate of each SDSS LRG.  As seen in
Figure~\ref{fig:f12}, the SFHs of the SDSS LRGs show good consistenty
with those of the mock LRGs, except that the SDSS LRGs have some
recent star formation, on the order of less than a few percent per
Gyr, at a lookback time $\la 6$\,Gyr. The later star formation found
in the SDSS LRGs may suggest that the feedback processes are not as
efficient as those adopted in the SAM model by \citet{Guo11} to
suppress star formation. Because of this later star formation, the mean
ages of the quiescent LRGs at a lower redshift, which were obtained from the full
spectrum fitting, may be systematically younger than those at a higher
redshift, even after correcting the difference in the lookback time.
It is, therefore, necessary to correct this systematical error to get a more accurate estimation of the Hubble constant (and other
cosmological parameters) by using the differential age method. 

\section{Uncertainties in estimating the Hubble constant by the
differential age method}
\label{sec:discussion}
%
%{\color {red}
Estimates of the Hubble constant, obtained from the differential age
method, may suffer some uncertainties caused by the later star
formation in LRGs and the systematic errors in the age estimates of
LRGs by using the stellar population synthesis technique.

We first estimate the systematical errors caused by the later star
formation in LRGs.  As shown in Figure~\ref{fig:f7}, the later star
formation rate is on the percentage per Gyr level  at $t \la 6$\,Gyr.
For illustration purposes, we assume that the mean later star formation
rate is $~2\%$ per Gyr at $t \la 6$\,Gyr. With this assumption, the
mean SFHs of LRGs at any redshift bin from $z=0$ to $0.4$ (with a
bin size of $\delta z=0.02$) can be obtained because the mean SFHs at
$t> 6$\,Gyr of those galaxies are given by the mean SFHs obtained for
the LRG samples given above. Consequently, the mean lifetime of those LRGs at
redshift $z$ can be estimated by
\begin{equation}
\left< t_z \right> = \sum_{j} \mu'_j (t_j + \delta t) + 0.02\Delta T.
\label{eq:tz}
\end{equation} 
Here $\delta t = \int^{0.4}_z \left| \frac{dt}{dz} \right| dz$,
$\Delta T = \int^{0.4}_z \left| \frac{dt}{dz} \right| dz$/Gyr,
$\mu'_j$ is the renormalized fraction of the $j$-th SSP, as shown in Figure~\ref{fig:f7}, and shifted to $z = 0.4$, the age of the
$j$-th SSP is replaced by $t_j+ \delta t$,  $\mu'_j = \mu_j/(\sum_j
\mu_j+0.02 \Delta T)$, and the summation $\sum_j \mu_j$ only covers
those SSPs with $t_j+ \delta t>0$.  To generate mock
observations, we set the Hubble constant as $H_0= 73 \kms{\rm
Mpc}^{-1}$ and the fraction of matter to the critical density as
$\Omega_{\rm m}= 0.25$.  According to the above procedures, we obtain
the mean age of the mock LRGs that is located at each redshift bin from $z=0$
to $0.4$ with a redshift interval of $\delta z=0.02$ , respectively,
and consequently we obtain the age-redshift relation. We then directly fit the
mock age-redshift relation using
\begin{equation}
t_{\age}=t_{\Univ}-t_{\form}.
\label{eq:tage}
\end{equation}
Here $t_{\age}$ is the mean age of the mock LRGs, $t_{\form}$ is the
mean formation time of those mock LRGs and assumed to be a constant
for each sample, and $t_{\Univ}$ is the cosmic age 
\begin{equation}
t_{\Univ}= \frac{1}{H_0} \int_{z}^{\infty} \frac{dz}{(1+z)
\sqrt{\Omega_{\rm m}(1+z)^3 + \Omega_{\Lambda}}},
\label{eq:tuniv}
\end{equation} where $\Omega_{\Lambda}$ is the fraction of the
cosmological constant to the critical density at the present time (see
details in \citealt{Liu12}).

Adopting the standard $\chi^2$ statistics and assuming a flat
universe, i.e., $\Omega_{\rm m} + \Omega_{\Lambda}=1$, we obtain the
best fit value of the Hubble constant as $H_0=92^{+1}_{-1}$,
$94^{+1}_{-1} $, $93^{+1}_{-1}$, and $93^{+1}_{-1} \kmsmpc$ from the
mock age-redshift relations obtained by using the mean SFHs, which were obtained
for the sub-sample of SDSS LRGs in the velocity dispersion ranges of
$210-240$, $240-270$, $270-300$, and $>300\kms$, respectively.
Apparently, a later star formation, at a rate of two percent stellar
mass per Gyr, can lead to an overestimation of the Hubble constant by
$\sim 28\%$, which is consistent with that argued in \citet[][see
discussions in Section 5.3 therein]{Liu12}. We note that, if the later
star formation rate is smaller, e.g., one percent in stellar mass per
Gyr, the $H_0$ value obtained from the mock age-redshift relation
would be correspondingly overestimated by $16\%$.

The mean age of quiescent LRGs may also not be accurate because of the
possible systematic uncertainty in the populations synthesis tool. For
example, the mean age of LRGs   is probably
overestimated by $0.09$\,dex in the logarithmic scale by using  STARLIGHT, as discussed in
Section~\ref{sec:compsammodel}). Therefore, it is important to
investigate the systematic error on the estimation of the Hubble
constant. Ignoring the later star formation in quiescent LRGs and
assuming a systematic overestimation of $0.09$\,dex in the age
measurements, we can also obtain the mock age-redshift relations by
procedures similar to those above, except that Equation~(\ref{eq:tz})
is replaced by
%the mean age of ``quiescent LRGs'' may be estimated as
%
\begin{equation}
\left< t_z \right> = \sum_{j} \mu'_j  (\delta t+ 10^{-0.09} t_j)
10^{0.09}.
\label{eq:crt}
\end{equation}
We find the best-fit $H_0= 63 \pm 1 $, $62 \pm 1$, $62 \pm 1 $, and
$61 \pm 1 \kmsmpc$ by fitting the mock age-redshift relations obtained
by using the mean SFHs for the SDSS LRGs in the $\sigma_v$ ranges of
$210$-$240$, $240$-$270$, $270$-$300$, and $>300\kms$, respectively.
Apparently, the systematic errors in the age estimations from using
$\textrm{STARLIGHT}$ can lead to an underestimation of the Hubble
constant of $\sim 16\%$. 

%If the systematic error in the age estimates is in linear scale, then
%it does not lead to any error in the measurements of the Hubble
%constant as this error can be absorbed into the formation time
%$t_{\rm U}$ in equation~(\ref{eq:tage}).

Considering both the systematic errors in the age estimates
($0.09$\,dex) and the late star formation in LRGs (2 percent of
stellar mass per Gyr), we can also obtain the mock age-redshift
relation by using the SFHs obtained for each subsample. We find that
the best-fit $H_0=74 \pm 2 $, $73 \pm 2$, $74 \pm 2$, and $73 \pm 2
\kmsmpc$, respectively. In this case, the uncertainty in the
estimation of the Hubble constant introduced by the later star
formation is largely canceled by the uncertainty introduced by the systematic
errors in the age estimates.

In summary, the later star formation in LRGs and the systematic errors
in the LRG age estimates lead to substantial uncertainty in the
estimation of the Hubble constant that is obtained from the differential age
method. However, this error can be corrected provided that the mean
SFH of the sample galaxies and the systematic error
in the age estimation are identified by using a specific stellar population synthesis
tool, such as STARLIGHT. We conclude that the Hubble
constant can be measured with high accuracy through the age--redshift
relation, taking    the effect of both
the later star formation and the systematic error in the age
estimates into careful consideration.

\section{Conclusions}
\label{sec:conclusion}

 Luminous red galaxies (LRGs) are believed to be passively evolving and
host the oldest stellar population, and as such,  can be used as cosmic
chronometers to measure the Hubble parameters at different redshifts.
Different LRGs may have different mass and locate in different
environments. The mass and environmental effects on
the LRG age properties,  if significant,  may limit the use of the LRGs as cosmic
chronometers. In this paper, we have investigated the environmental and mass
effects on the formation and evolution of quiescent SDSS LRGs through
the stellar population synthesis technique and the full spectrum
fitting method and have checked the importance of the environmental effects
on the application of LRGs as cosmic chronometers. We find that (1)
the ages of quiescent LRGs, whether they are BCGs of clusters, member
galaxies of clusters (MGs), or field galaxies (FGs),  correlate
weakly with their velocity dispersions, i.e., the larger the velocity
dispersion, the slightly older the ages; (2) the age distributions and 
the mean SFHs of quiescent LRGs with similar velocity dispersions in
different environments do not show significant differences;
(3) the SFHs of the SDSS LRGs are well consistent with those of the
mock LRGs that result from a semi-analytical model of galaxy formation,
except that the SDSS LRGs have some recent star formation (at a
lookback time $\la 6$\,Gyr of a few percent per Gyr; and
(4) the systematic errors in the age estimates of LRGs and the later
star formation in LRGs all lead to some errors in estimating the
Hubble constant through the differential age method.  We find that the
late star formation in quiescent SDSS LRGs leads to a systematical
overestimation of the Hubble constant; although the systematic errors in
the STARLIGHT-derived ages of LRGs may lead to an underestimation of
the Hubble constant.  However, both effects may be corrected by a
careful study of the mean SFH of those LRGs and by calibrating the
STARLIGHT-derived ages with those obtained independently using other
methods. 

In summary,  environmental effects don not play a  significant role in the age
estimates of quiescent LRGs;  as a population,
the quiescent LRGs can be securely used as cosmic chronometers and the Hubble constant
can be measured with high precision by the differential age method, as
performed by \citet{Jim03}, \citet{Simon05}, \citet{Stern10},
\citet{Moresco12}, and \citet{Liu12}.

\begin{acknowledgements} 
We are grateful to the anonymous referee for constructive suggestion.
Gaochao Liu thanks Zhonglue Wen for providing the BCG and LRG
catalog and Xianmin Meng for helpful discussions on using
\texttt{STARLIGHT}.  This work is partly supported by NSFC grant
support under Nos.~11303020, 11073024, 11203033, 11373031, 11390372,
U1331113, U1231123, U1331202, the Strategic Priority Research Program
``The Emergence of Cosmological Structures'' of the Chinese Academy of
Sciences, grant No. XDB09000000, and the CAS grant KJCX2-EW-W01.

 Funding for SDSS-III has been provided by
the Alfred P. Sloan Foundation, the Participating Institutions,
the National Science Foundation, and the USDepartment of
Energy. The SDSS-III Web site is http://www.sdss3.org/.
SDSS-III is managed by the Astrophysical Research Consortium
for the Participating Institutions of the SDSS-III Collaboration
including the University of Arizona, the Brazilian
Participation Group, Brookhaven National Laboratory, University
of Cambridge, University of Florida, the French Participation
Group, the German Participation Group, the Instituto
de Astrofisica de Canarias, the Michigan State/Notre
Dame/JINA Participation Group, Johns Hopkins University,
Lawrence Berkeley National Laboratory, Max Planck Institute
for Astrophysics, New Mexico State University, New
York University, Ohio State University, Pennsylvania State
University, University of Portsmouth, Princeton University,
the Spanish Participation Group, University of Tokyo, University
of Utah, Vanderbilt University, University of Virginia,
University of Washington, and Yale University.
\end{acknowledgements}

%-------------------------------------------------------------------

\appendix

\section{Distribution of age and mass fraction }
 
\begin{figure*}
\centering
\includegraphics[scale=0.7]{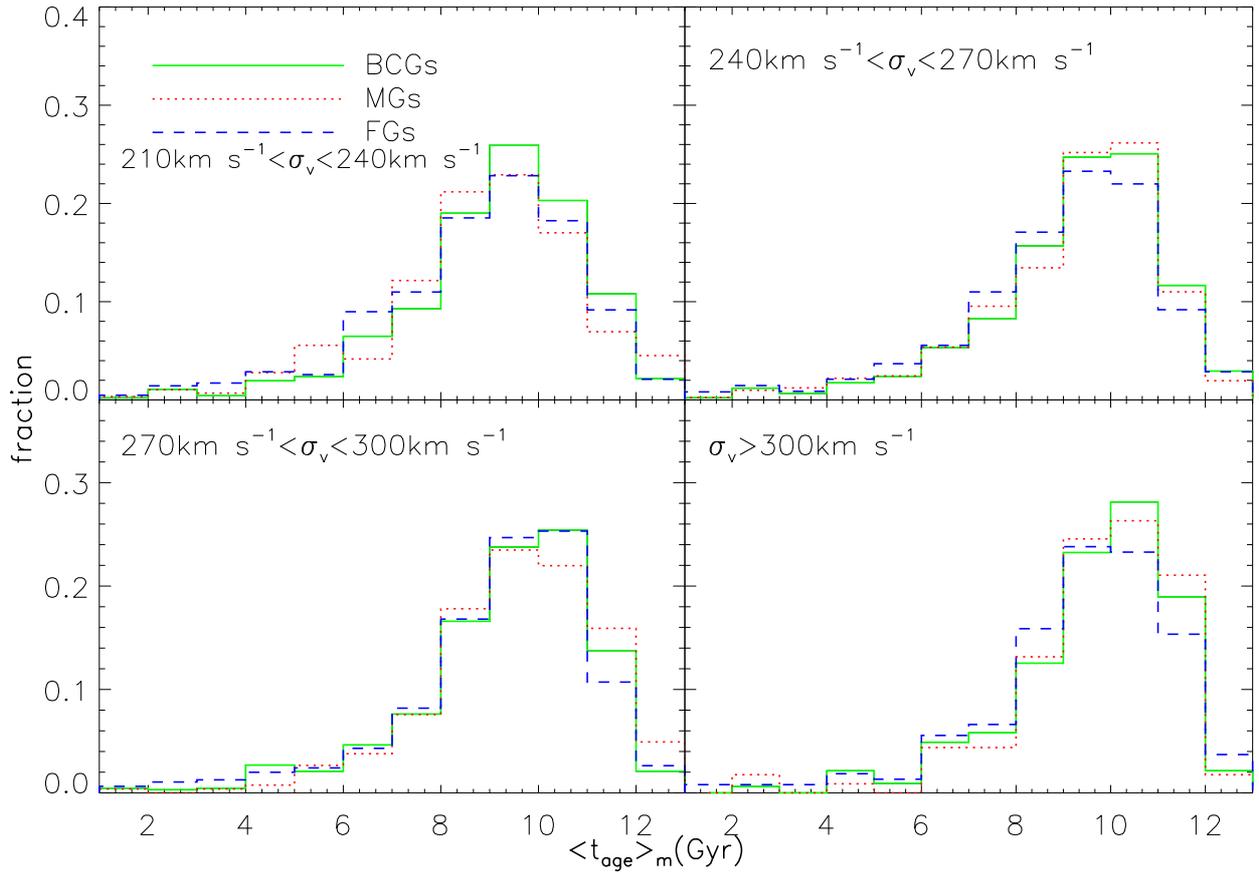}
\caption{Age distributions of BCGs (green solid line), MGs (red dotted
line), and FGs (blue dashed line).  }
\label{fig:appendix1} 
\end{figure*}

\begin{figure*}
%\begin{figure}
\centering
\includegraphics[scale=0.7]{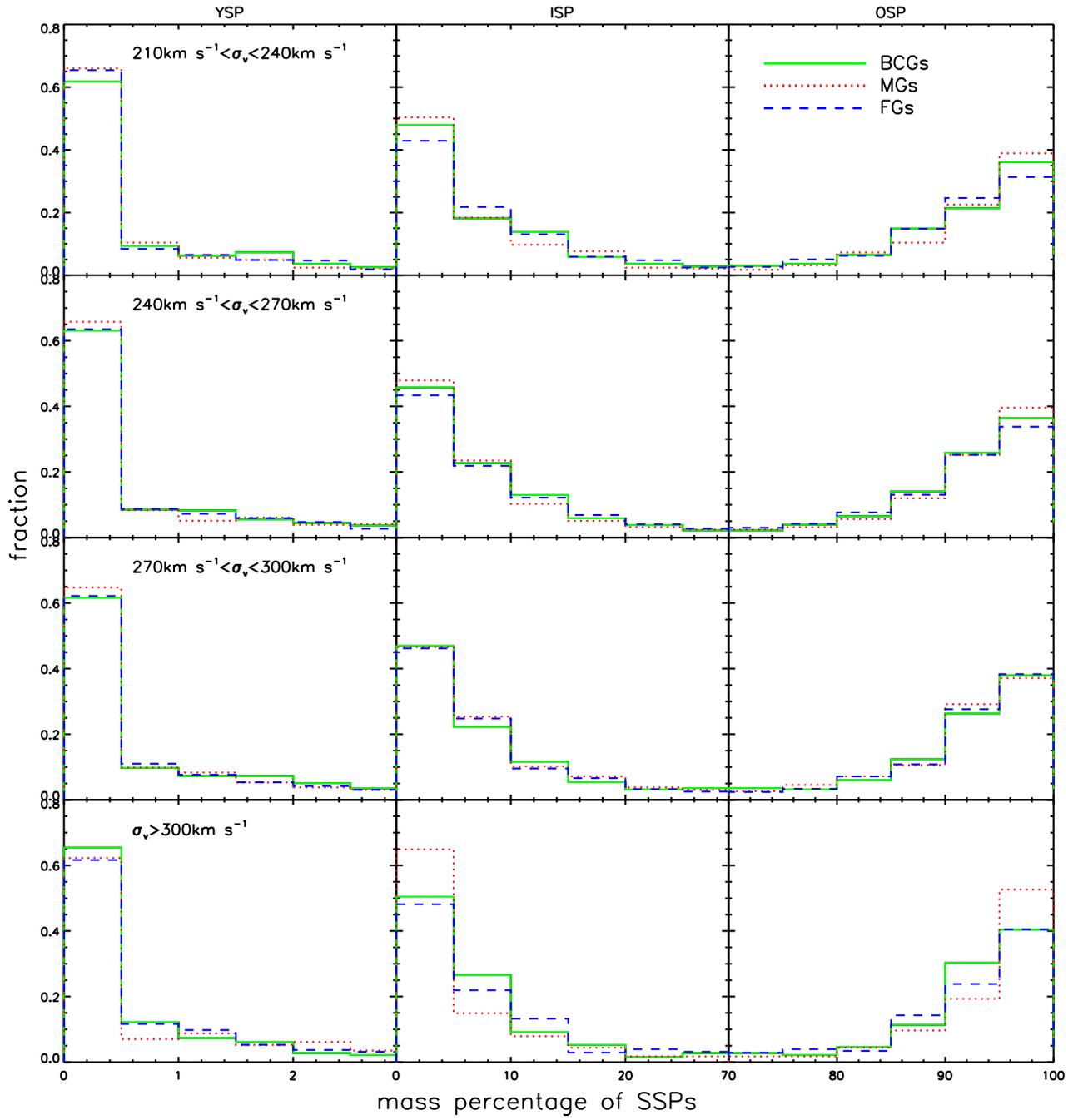}
\caption{Distributions of the mass fractions of different SSPs of LRGs
in different subsamples.  Histograms shown by the solid (green),
dotted (red), and dashed (blue) lines represent the distribution for
BCGs, MGs, and FGs, respectively.  }
\label{fig:appendix2} 
%\end{figure}
\end{figure*}

In the main text, Figure ~\ref{fig:f5} shows the age distributions of
BCGs, MGs, and FGs with $\sigma_v$ in the range from $270\kms$ to
$300\kms$,  and Figure~\ref{fig:f8} shows the distribution of the
mass fractions of YSP, ISP, and OSPs in the BCGs, MGs, and FGs with
the same velocity dispersion bin. For completeness, we also show here the age
distributions and the distributions of the mass fractions of BCGs, MGs, and FGs
for other velocity dispersion bins in Figure \ref{fig:appendix1} and
Figure \ref{fig:appendix2} , respectively.

\end{document}